\documentclass[showpacs,floatfix,preprint]{revtex4-1}
\usepackage{graphicx,psfrag,amsmath,amssymb,amsfonts,latexsym,color,epsf,graphpap,dcolumn}
\usepackage{caption}
\usepackage{subcaption}

\begin{document}
\title{Dynamical Casimir effect for semitransparent mirrors}
\author{C.~D.~Fosco$^{a,b}$, A Giraldo$^{b}$ and F.~D.~Mazzitelli$^{a,b}$}
\affiliation{$^a$Centro At\'omico Bariloche,  CONICET,
Comisi\'on Nacional de Energ\'\i a At\'omica, R8402AGP Bariloche, Argentina.\\
$^b$Instituto Balseiro, Universidad Nacional de Cuyo, R8402AGP Bariloche, Argentina }
\date{today}
\begin{abstract} 
\noindent We study the Dynamical Casimir Effect resulting from the
oscillatory motion of either one or two flat semitransparent mirrors,
coupled to a quantum real and massless scalar field. Our approach is based on
a perturbative evaluation, in the coupling between mirrors and field, of
the corresponding effective action, which is used to compute the particle creation rate.
The amplitude of the oscillation is not necessarily small. 
We first obtain results for a single mirror, both for non-relativistic and
relativistic motions, showing that only for the latter the effects may be
significant.  For two mirrors, on the other hand, we show that there are 
interesting interference effects, and that in some particular cases the results 
differ from those obtained assuming small amplitudes,  
already for non-relativistic motions. 
\end{abstract}
\maketitle
\section{Introduction}\label{sec:intro}

Dynamical Casimir effect (DCE), or motion induced radiation,  refers to a
plethora of phenomena in which real particles are created from the quantum
vacuum due to the presence of external, time-dependent conditions.  The
creation of particles in a one dimensional cavity with a moving perfect
mirror was first studied by Moore~\cite{Moore}, and subsequently by Fulling
and Davies~\cite{FulDav}, as a toy model of black hole evaporation. Over
the years, the DCE has received increasing attention,  and became a
relevant topic in studies on cavity quantum electrodynamics and cavity
optomechanics, superconducting waveguides with time dependent boundary
conditions, refractive index perturbations in optical fibers,
quantum friction, etc,  in addition to analogue gravity models. For some
recent reviews see~\cite{dodonov, dalvit, schutzhold,nation}.

In this work we evaluate the particle creation rate for a system which
consists of either one or two flat, infinite, parallel semitransparent
mirrors, undergoing oscillatory motion.  Non-perfect moving mirrors have
been considered long time ago in Ref.~\cite{Barton1}, where the authors
studied the quantum radiation from a dispersive mirror moving
non-relativistically in $1+1$ dimensions. Later on, more general models have
been considered by several authors~\cite{gensemi}. 

Our approach here relies on a main assumption, namely, that of the mirrors
being semitransparent, what justifies our use of a perturbative expansion
in the strength of the coupling between each mirror and the quantum field.
This approach is the dynamical counterpart of the perturbative calculations 
of the static Casimir force for dilute dielectric  bodies \cite{Golestanian, Milton},  in which the 
small parameter is $\epsilon -1$, where $\epsilon$ is the permittivity.

Since the amplitude of the oscillatory motion(s) is not assumed to be
necessarily small, our results will be non-perturbative in that amplitude,
and therefore our approach may be regarded as complementary to others which
are non-perturbative in the mirror-field couplings but restricted to small 
amplitudes and non-relativistic motion. We will consider a simplified model involving a vacuum real scalar field,
with the mirrors described by means of $\delta$-potentials, and will present calculations
up to second order in the coupling constants.  The results could be generalized to
more realistic models involving the electromagnetic field and to higher perturbative
orders. In particular, we have in mind 
situations in which there is  particle creation due to a varying refractive index perturbation
$n(t,\mathbf x)=n_0+\delta n(t,\mathbf x)$ with $\delta n \ll 1$ \cite{schutzhold,rip}.

The model considered in this paper has been first analyzed in the context of the DCE in 
Ref.\cite{Barton1}, in $1+1$ dimensions.  It was pointed out there that, due to infrared divergences,
an approach perturbative in the coupling constant is not possible.  However, these divergences are 
typical for two dimensional massless fields. As we will
see, they disappear in higher dimensions and the perturbative calculations are perfectly 
well defined (to our knowledge, this point has not been explored before).  
Generalizations of the $\delta$-potential models have been considered
more recently in the context of optomechanics \cite{Hu}.

This paper is organized as follows: in Section~\ref{sec:thesys} we describe
the system that we consider subsequently, and introduce our conventions and
notation. Then we consider the in-out effective
action, presenting the corresponding weak-coupling expansion. 
In Section III we compute the imaginary part of the effective action for the case
of an oscillating single mirror.  
Using  the Jacobi-Anger expansion it is possible to compute
the imaginary part
of the effective action for oscillatory 
motions  of arbitrary amplitude, including relativistic corrections.
In Section IV we consider the case of two oscillating mirrors. We discuss interference 
effects and compare the results with those coming from a small amplitude approximation.
Section V contains the conclusions of our work.

\section{Classical model and in-out effective action}\label{sec:thesys}

\subsection{The classical action}

We follow the functional integral formalism, whereby the system is
defined in terms of its (real time) action ${\mathcal S}$, for a real
scalar field $\varphi$ in $D \equiv d+1$ dimensions. The action also
depends on the configuration of the mirror (or mirrors), which play the
role of `external fields' here. In the examples that we shall consider,
they are assumed to be infinite and parallel planes \cite{footnote}, 
it is sufficient to give just one function of time to determine
the position of each mirror.  Furthermore, we assume ${\mathcal S}$ to have
the structure:
\begin{equation}
{\mathcal S}\;=\; {\mathcal S}_0 \,+\, {\mathcal
S}_I \;,
\end{equation}
where ${\mathcal S}_0$ denotes the free real scalar field action: 
\begin{equation}
{\mathcal S}_0(\varphi) \;=\; \frac{1}{2} \int  d^Dx \; 
\partial_\mu\varphi \partial^\mu\varphi  \;,
\end{equation}
the part of the total action which is independent of the configurations of
the mirrors.
The ${\mathcal S}_I$ term accounts, on the other hand, for the interaction
between the field and the mirror(s). For a single, flat, zero-width
mirror, moving along the normal direction to its plane, its instantaneous
position may be completely determined by an equation with the form:
\begin{equation}
	x^d \;=\; q(x^0) \;, 
\end{equation}
where $x^d$ denotes the coordinate normal to the plane.  The class of mirrors
considered in this work has, therefore, translation invariance along the
spatial `parallel' coordinates ${\mathbf x}_\parallel \equiv (x^1,\ldots,
x^{d-1})$.  Besides, the coordinates which are relevant to describe the motion
will also be denoted (irrespective to the value of $d$) as $z\equiv x^d$, and
$t \equiv x^0$, and we will assume its action ${\mathcal S}_I$ to be given by:
\begin{equation}\label{eq:single}
{\mathcal S}_I \;=\; - \frac{\lambda}{2} \, \int d^{d+1}x  \,
\gamma^{-1}(t) \,\delta[z - q(t)]\; \varphi^2(x) 
\end{equation}
where $\lambda$ is a constant that determines the strength of the coupling,
and $\gamma(t)$ denotes the Lorentz factor: $\gamma(t) \equiv 1/\sqrt{1
- \dot{q}^2(t)}$, which is only relevant to the relativistic-motion case
(in our conventions, the speed of light $c \equiv 1$).

When considering two mirrors, denoted by $L$ and $R$, no {\em direct\/}
coupling is assumed to exist between them, aside from the indirect one
which will result from the mediation of the scalar field.  Thus the total
interaction action becomes the sum of the corresponding terms ${\mathcal
S}={\mathcal S}_L+{\mathcal S}_R$, where 
\begin{equation}
{\mathcal S}_{L,R} \;=\; - \frac{\lambda_{L,R}}{2} \, \int d^Dx  \,
\gamma^{-1}_{L,R}(t) \,\delta[z - q_{L,R}(t)]\; \varphi^2(x)\;,
\end{equation}
with $\gamma_{L,R}(t) \equiv 1/\sqrt{1 - \dot{q}_{L,R}^2(t)}$,
where the functions $q_L$ and $q_R$ define the motion of the respective
mirror.  

The action considered here can be thought as a toy model for the interaction of a non-perfect mirror 
with the electromagnetic field and, with some modifications, for a situation in which a
refractive index perturbation concentrated on a plane travels along a trajectory given by $x^d=q(x^0)$.

\subsection{The effective action}\label{sec:effective}
The (in-out) effective action, is a functional of the functions that
determine the instantaneous position the mirrors, and is simply related to
the vacuum persistence amplitude, namely,
\begin{equation}
	e^{i \Gamma } \;=\; \int {\mathcal D}\varphi \; e^{i{\mathcal
	S}(\varphi)}=\langle0_{out}\vert 0_{in}\rangle_{q(x^0)} \;.
\end{equation}
The probability $\mathcal P$ of pair creation of real particles associated
to the vacuum field,  during the whole motion of the mirror(s), is given by
\begin{equation}
\label{prob}
e^{-2 {\rm Im} \left[\Gamma \right]} \;=\; 1-{\mathcal P}\, . 
\end{equation}
We note first that $\Gamma$ can be split as $\Gamma = \Gamma_0 + \Gamma_I$,
where $\Gamma_0$ is the effective action corresponding to the free action
${\mathcal S}_0$, and therefore will be discarded. On the other hand,
\begin{equation}
	e^{i \Gamma_I} \;=\; \langle e^{i {\mathcal S}_I(\varphi)} \rangle
	\;,
\end{equation}
where the average symbol of a given functional of the vacuum field is
understood in the functional sense, with ${\mathcal S}_0$  defining a
(complex) Gaussian weight: 
\begin{equation}
\langle \; \ldots \; \rangle \;\equiv\; \frac{\int {\mathcal D}\varphi \ldots
e^{i{\mathcal S}_0(\varphi)}}{\int {\mathcal D}\varphi \; e^{i {\mathcal
S}_0(\varphi)}} \;.
\end{equation}

For weak coupling, we use an expansion in cummulants, which proceeds as
follows: we assume the strength of the ${\mathcal S}_I$ term, controlled by
the value of the $\lambda$ coefficients, is such that we may expand
$\Gamma$ in powers of that term.  $\Gamma_I$ may then be expanded in powers
of ${\mathcal S}_I$; denoting by $\Gamma_I^{(k)}$ the $k^{th}$-order term
in that expansion, we see that:
\begin{equation}
	\Gamma_I \;=\; \Gamma_I^{(1)} \,+\,\Gamma_I^{(2)}\,+\,\ldots
	\,+\,\Gamma_I^{(k)} \,+\,\ldots 
\end{equation}
where
\begin{equation}\label{GammaI}
\Gamma_I^{(1)} = \langle {\mathcal S}_I \rangle \;\;,\;\;
	\Gamma_I^{(2)}=\frac{i}{2} \langle ({\mathcal S}_I - \langle
	{\mathcal S}_I\rangle)^2 \rangle\;, \ldots,\;
	\Gamma_I^{(n)}=\frac{i^{n-1}}{n!} \langle ({\mathcal S}_I)^n
	\rangle_c \;\;,\ldots 
\end{equation}
where the subscript $c$ denotes the {\em connected\/} part of the
Feynman diagrams (resulting from the application of Wick's theorem to the
calculation of the Gaussian averages).

In what follows, we deal with the explicit evaluation of the imaginary part
of the effective action for either one or two mirrors (as mentioned previously, this quantity
is related to the probability of particle creation Eq.\eqref{prob}). Our focus shall be
in the $D=4$ case, although we will also comment on some particular cases
where different values of $D$ produce qualitatively different results.

\section{A single mirror}
The first-order term in Eq.\eqref{GammaI}
leads to
\begin{equation}
	\Gamma_I^{(1)} \;=\;- \frac{\lambda}{2} \, 
	\int d^4x \,\gamma^{-1}(t) \, \delta(z - q(t) ) \,   
	\langle \varphi(x) \varphi(x) \rangle 
\end{equation}
which may be interpreted as an infinite  renormalization for the
mirror, regarded as a particle moving in one spatial dimension
(corresponding to the $z$ coordinate). Thus, $\Gamma_I^{(1)}$ may be
written as follows:
\begin{equation}
\Gamma_I^{(1)} \; =\;- m_\Lambda \, \int d\tau  \;,  
\end{equation}
where $\tau$ denotes the proper time corresponding to the trajectory
defined by $q(t)$, and the mass $m_\Lambda$, regularized by means of an
UV cutoff $\Lambda$, is given by:
\begin{align}
m_\Lambda &=\; \frac{\lambda}{2} \,L^{d-1}\,  \langle
	\varphi^2(x)\rangle_\Lambda \nonumber\\ 
 &=\; \frac{\lambda}{2} \,L^{d-1}\, \int_\Lambda \frac{d^Dp}{(2\pi)^D} \,
  \frac{1}{p^2} \,=\, \xi \,\lambda \,(L \Lambda)^{d-1} \;,
\end{align}
where $L$ has the dimensions of a length, and $L^{d-1}$ is the total
`volume' of the mirror, as measured along the $d-1$ spatial coordinates
which are parallel to its surface. $\xi$ is a dimensionless,
regularization-dependent constant.

For the second-order term, we see that:
\begin{align}\label{eq:gi2rsing}
\Gamma_I^{(2)} &=\; \frac{i\lambda^2}{4} \; 
	\int d^Dx \int d^Dx' \; \gamma^{-1}(t) \delta(z - q(t)) \,
	\gamma^{-1}(t') \delta(z' -
	q(t')) \big( \langle \varphi(x) \varphi(x') \rangle \big)^2
	\nonumber\\
& \equiv \; \frac{1}{2} \; 
	\int d^Dx \int d^Dx' \; \gamma^{-1}(t) \, \delta(z - q(t)) \; \Pi^{(2)}(x,x') \;
	\gamma^{-1}(t') \, \delta(z' - q(t')) \;.
\end{align}
where
\begin{equation}
	\Pi^{(2)}(x,x') \;=\;\frac{i\lambda^2}{2} \; \Big( G(x,x') \Big)^2 \,
\end{equation}
with  the free Feynman propagator $G$ for the scalar field  given by:
\begin{align}
G(x,x') &=\, G(x-x') \,=\, \langle \varphi(x) \varphi(x') \rangle
	\nonumber\\
&=\, \int \frac{d^Dp}{(2\pi)^D} \, e^{-i p \cdot (x-x')} \,\widetilde{G}(p)
\;,\;\;\; \widetilde{G}(p) \,\equiv\, \frac{1}{p^2 + i 0^+} \;.  
\end{align}
Thus, in Fourier space, Eq.(\ref{eq:gi2rsing}) becomes:
\begin{equation}\label{eq:gi21rsing}
	\Gamma_I^{(2)} \;=\; \frac{1}{2}\, \int \frac{d^Dk}{(2\pi)^D}\; 
	\widetilde{\Pi}^{(2)}(k) \, \big| \tilde{F}_\gamma(k)\big|^2 \;,
\end{equation}
where 
\begin{equation}
	\widetilde{\Pi}^{(2)}(k) \;=\; \frac{i \lambda^2}{2} \; \int
\frac{d^Dp}{(2\pi)^D} \; \widetilde{G}(p) \, \widetilde{G}(p-k)
\end{equation}
and:
\begin{align}
	& \tilde{F}_\gamma(k) \;=\; \int d^Dx \, e^{i k \cdot x} \,
	\gamma^{-1}(x^0) \, \delta[x^d - q(x^0)] \,=\,
	(2\pi)^{d-1} \, \delta^{d-1}({\mathbf k}_\parallel) \, 
\tilde{f}_\gamma(k^0,k^d) \label{Fgamma}\\
& \tilde{f}_\gamma(k^0,k^d) \;=\; \int_{-\infty}^{+\infty} dt \,
\gamma^{-1}(t) \, e^{i k^0 t} \, e^{-i k^d q(t)} \;.
\label{fgamma}\end{align}

In this perturbative approach, the probability of creation of a pair of particles is
\begin{equation}
{\mathcal P}\simeq
2{\rm Im}\left[\Gamma_I^{(2)}\right ] = \int \frac{d^Dk}{(2\pi)^D}\; 
	{\rm Im}\left[\widetilde{\Pi}^{(2)}(k)\right] \, \big| \tilde{F}_\gamma(k)\big|^2 \;,
\end{equation}
and therefore the total energy $\mathcal E$ of the created particles reads
\begin{equation}\label{energy}
{\mathcal E}\simeq \int \frac{d^Dk}{(2\pi)^D}\; 
	{\rm Im}\left[\widetilde{\Pi}^{(2)}(k)\right] \, \big| \tilde{F}_\gamma(k)\big|^2 \vert k_0\vert \;.
\end{equation}

\subsection{Low-velocity motion}

We now assume the explicit form of $q(t)$, the mirror's oscillatory motion, to
be harmonic:
\begin{equation}\label{eq:mot}
	q(t) \;=\; \epsilon \; \cos \Omega t \;\;,
\end{equation}
We will first consider a non-relativistic motion, with $\epsilon\,\Omega\ll 1$ and therefore $\gamma\simeq 1$ (in which we will
denote $\tilde f_\gamma\equiv \tilde f$ ), and then compute corrections in powers of the maximum velocity $v=\epsilon\;\Omega$.

The usual approaches to this problem  consider  small amplitudes for the oscillation, then  the exponential in Eq.\eqref{fgamma}
is expanded in powers of $\epsilon$
\begin{equation}
e^{-i k^d q(t)}\simeq 1 - i k^d \; \epsilon \; \cos \Omega t -\frac{1}{2} ( k^d \; \epsilon \; \cos \Omega t )^2 + \cdots\, .
\end{equation}
Instead of doing this, we use the Jacobi-Anger expansion and get:
\begin{equation}
	\tilde{f}(k^0,k^d) \;=\; 2\pi \, 
	\sum_{n=-\infty}^{+\infty} \,i^n \,  J_n(k^d \epsilon) \,
	\delta(k^0 - n \Omega) \;,
\end{equation}
where $J_n$ denotes a Bessel function of the first kind. This expression is valid for any amplitude value. The effective action
$\Gamma_I^{(2)}$ becomes extensive in the volume of the mirror, and
proportional to the extent $T$ of the time coordinate. Thus, defining
$\gamma_I^{(2)} \equiv \Gamma_I^{(2)}/(T L^{d-1})$, which has the
dimensions of an energy per unit of $d-1$-dimensional volume:
\begin{equation}\label{eq:gi22sing}
	\gamma_I^{(2)} \;=\; \frac{1}{4\pi}\, \sum_{n=-\infty}^{+\infty} \, 
	\int_{-\infty}^{+\infty} dk^d\;
	\big[\widetilde{\Pi}^{(2)}(k)\big]\Big|_{k^0 = n
	\Omega,\,{\mathbf k}_\parallel = {\mathbf 0}} \; [J_n(k^d
	\epsilon)]^2 \;.
\end{equation}
Finally, since we are interested in the imaginary part of the effective
action, we get, to this order:
\begin{equation}\label{eq:gi23sing}
	{\rm Im}\Big[\gamma_I^{(2)}\Big] \;=\; \frac{1}{4\pi}\, \sum_{n=-\infty}^{+\infty} \, 
	\int_{-\infty}^{+\infty} dk^d\;
	{\rm Im}\big[\widetilde{\Pi}^{(2)}(k)\big]\Big|_{k^0 = n
	\Omega,\,{\mathbf k}_\parallel = {\mathbf 0}} \; [J_n(k^d
	\epsilon)]^2 \;.
\end{equation}
In other words, the imaginary part of the effective action is determined by
the imaginary part of $\widetilde{\Pi}^{(2)}$, at some special points in
momentum space, which depend on the properties of the mirror's oscillation.

To proceed, we need a more explicit expression for
$\widetilde{\Pi}^{(2)}(k)$.  This object is, generally,  UV divergent, but
not its imaginary part, since the
divergences are (at most) polynomials in $k$, i.e., entire functions.

By using a Feynman parameter $\alpha$, the $p$ momentum integral can be
performed using dimensional regularization, the result being: 
\begin{equation}
	\widetilde{\Pi}^{(2)}(k) \;=\; - \frac{\lambda^2}{2} \, \frac{\Gamma(2-
	D/2)}{(4\pi)^{\frac{D}{2}}} \, \int_0^1 d\alpha \, \Big[- \alpha
	(1-\alpha) k^2 \Big]^{\frac{D-4}{2}} \;. 
\end{equation}
This is IR divergent when $D=2$, while for $D=3$ it is IR and UV finite,
its form being:
\begin{equation}
	\widetilde{\Pi}^{(2)}(k) \;=\; - \frac{\lambda^2}{16} \,
	(-k^2)^{-1/2} \;. 
\end{equation}
Thus,
\begin{equation}
	{\rm Im}\Big[\widetilde{\Pi}^{(2)}(k)\Big] \;=\; \frac{\lambda^2}{16}
	\,\theta\big(|k^0|-|{\mathbf k}|\big) \, 
	(k^2)^{-1/2} \;\;\; (D=3)\;. 
\end{equation}
In the particular case of $D=4$, we have a UV divergent term which, being a
constant, does not contribute to the imaginary part of the effective
action. Thus, from the minimally subtracted part:
\begin{equation}
	\widetilde{\Pi}^{(2)}(k) \;=\; \frac{\lambda^2}{32 \pi^2} \,
	\log(-k^2) \;. 
\end{equation}
we obtain
\begin{equation}
	{\rm Im}\Big[\widetilde{\Pi}^{(2)}(k)\Big] \;=\; \frac{\lambda^2}{32
	\pi^2} \,\theta\big(|k^0|-|{\mathbf k}|\big) \pi \;\;\;(D=4) \;, 
\end{equation}
where $\theta$ denotes Heaviside's step function.

We can then write down the explicit results for $D=3$ and $D=4$ to this
order:
\begin{align}\label{Im34}
{\rm Im}\Big[ \gamma_I^{(2)} \Big]_{D=3} &=\; \frac{\lambda^2}{16\pi} 
	\sum_{n=1}^{+\infty} \, \int_0^{n |\Omega|} dk^d\; 
	\frac{[J_n(k^d \epsilon)]^2}{\sqrt{(n \Omega)^2 - (k^d)^2}} \nonumber\\
{\rm Im}\Big[\gamma_I^{(2)}\Big]_{D=4} &=\; \frac{\lambda^2}{32 \pi^2}
	\sum_{n=1}^{+\infty} \, \int_0^{n |\Omega|} dk^d\; 
	[J_n(k^d \epsilon)]^2  \;.
\end{align}

Note that from Eq.~\eqref{energy}, one could also extract the total power
dissipated per unit area: one should insert a factor $2n\Omega$
into the series in Eq.~\eqref{Im34}, and divide the result by the total
time elapsed. 

Let us now compare our $D=4$ result in Eq.\eqref{Im34}, with
the result of a calculation perturbative in the amplitude.  To that end, we
expand the Bessel functions for small arguments; we see that the
leading contribution comes just from the $n=1$ term. Hence:
\begin{equation}\label{imgpert}
{\rm Im}\Big[\gamma_I^{(2)}\Big]_{D=4} =\; \frac{\lambda^2}{32 \pi^2}
	\sum_{n=1}^{+\infty} \, \int_0^{n |\Omega|} dk^d\;
	[J_n(k^d \epsilon)]^2\simeq \frac{\lambda^2\epsilon^2\vert\Omega\vert^3}{384 \pi^2}\, ,
\end{equation}
which is consistent with the results in Ref.~\cite{fosetal1} for the
specific motion of the mirror defined by Eq.~\eqref{eq:mot}. A numerical
evaluation of the series shows that the leading perturbative result is
highly accurate in the non-relativistic limit $\epsilon\, \Omega\leq 0.1$.
This is illustrated in Fig.~1.
\begin{figure}[!ht]
\centering
\includegraphics[scale=1.0]{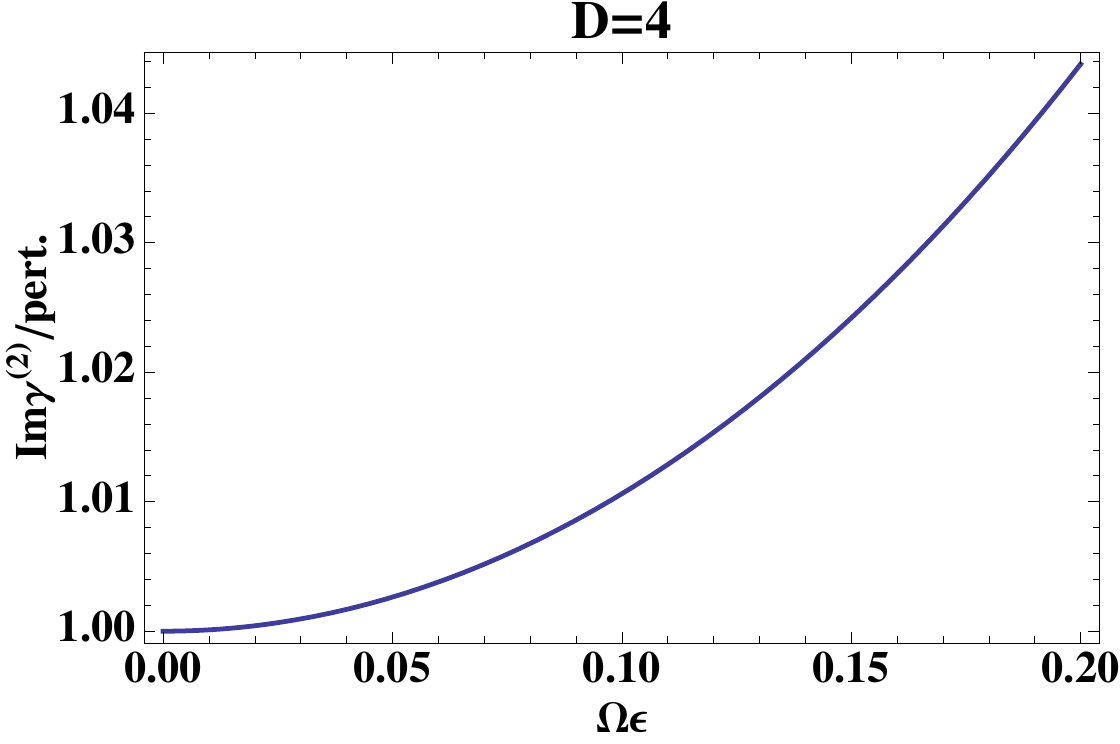} 
\caption{\label{fig1}Imaginary part of the effective action  for a single mirror, normalized by the perturbative result,  as a function of $\Omega\epsilon$. }
\end{figure}

We now compute the relativistic corrections to these results, expanding  the $\gamma^{-1}$ factor in $\tilde f_\gamma$ in powers of $v$. From Eq.\eqref{fgamma} we obtain
\begin{equation}
\tilde f_\gamma-\tilde f \equiv \Delta \tilde f=   -\frac{v^2}{2}  \int_{-\infty}^{+\infty} dt  \, e^{i k^0 t} \, e^{-i k^d q(t)}\sin^2\Omega t + \mathcal{O} (v^4)\; .
\end{equation}
Using again the Jacobi-Anger expansion we get
\begin{equation}
\Delta\tilde f = \frac{v^2\pi }{4}\sum_{n=-\infty}^\infty (-i)^n J_n(k^d \epsilon)
\left[\delta(k_0-(n+2)\Omega)+\delta(k_0-(n-2)\Omega)
- 2 \delta(k_0-n\Omega)\right]\, ,
\end{equation}
and therefore
\begin{equation}
\Delta {\rm Im}\Big[\gamma_I^{(2)}\Big]_{D=4} =\; \frac{\lambda^2 v^2}{128 \pi^2}
	\sum_{n=1}^{+\infty} \, \int_0^{n |\Omega|} dk^d\;J_n\left(J_{n+2}+J_{n-2}-2J_n\right)\vert_{k^d\epsilon}\, .
\label{eq:deltaim}
\end{equation}	

To compare this with the non-relativistic result, we write both in a way which shows the 
dependence in $v$ more explicitly:
\begin{equation}\label{imgeq}
{\rm Im}\Big[\gamma_I^{(2)}\Big]_{D=4} =\; \frac{\lambda^2 \Omega}{32 \pi^2} 
\, \frac{1}{v} \, 
\sum_{n=1}^{+\infty} \, \int_0^{n v} du\; [J_n(u)]^2
\end{equation}
and 
\begin{equation}\label{comp2}
\Delta {\rm Im}\Big[\gamma_I^{(2)}\Big]_{D=4} =\; \frac{\lambda^2 \Omega v}{32 \pi^2}
	 \, \sum_{n=1}^{+\infty} \, \int_0^{n v} du\;  J_n(u) \frac{d^2 J_n}{du^2}(u) \;,
\end{equation}	
where the latter has been obtained from Eq.\eqref{eq:deltaim} by using recurrence relations of the Bessel functions.

We made a numerical evaluation of these expressions, and found that
the series in Eqs.\eqref{imgeq} and \eqref{comp2} are dominated by the first terms, and are of the same order of magnitude. 
More concretely, we found that the relativistic correction can be fitted as 
\begin{equation}
\frac{\Delta {\rm Im}\Big[\gamma_I^{(2)}\Big]_{D=4}}{{\rm Im}\Big[\gamma_I^{(2)}\Big]_{D=4}} \simeq  \, 0.25\, v^2\, ,
\label{eq:dim/im}
\end{equation}
for $v<0.3$.

\subsection{Ultrarelativistic motion}

In order to obtain closed analytical expressions in the ultrarelativistic case,
we have found it convenient to consider, instead of a harmonic motion, the
oscillatory motion $q(t) = \epsilon \, \eta(t)$, where $\epsilon$
sets the amplitude of motion, and $\eta(t)$ oscillates between $-1$ and
$1$, depending linearly on $t$ in each half-period:
\begin{equation}\label{relmot}
\eta(t) \;=\; \left\{ 
\begin{array}{cc} 
1 - \Omega t \,,&\; 0 \leq  t < \frac{2}{\Omega} \\	
1 + \Omega ( t - \frac{4}{\Omega}) \,,&\;\frac{2}{\Omega} \leq t < \frac{4}{\Omega} \;.	
\end{array}
\right. 
\end{equation}
Note that this motion has infinite acceleration at the return points.
Physically, this is an idealization of a smooth trajectory in which the
change of velocity at the return points takes place during a small 
finite time interval $\Delta t\ll 1/\Omega$, and therefore the acceleration
at those points is of the order $\vert a\vert=2\epsilon\Omega/\Delta t$.
We shall keep in mind that this finite returning time $\Delta t$  will act as a ``cutoff'', 
a fact that will reflect itself when encountering an UV
divergence below. Note that this cutoff is a parameter of the motion, not 
an artifact of the calculation. A similar motion (but with a single half-period), has been considered before
by Moore \cite{Moore} and Fulling and Davies \cite{FulDav}, in $1+1$ dimensions.

The main reason to introduce the idealization above is a practical one: it
simplifies the calculation. Indeed, (neglecting the contribution of this small interval) the velocity in Eq.~\eqref{relmot}
always has modulus $v = \epsilon \Omega$, and the Lorentz factor becomes
time-independent. 
Thus,
\begin{equation}
	\tilde{f}_\gamma(k^0,k^d) \;=\; \sqrt{1 -(\epsilon \Omega)^2} 
		\, \int_{-\infty}^{+\infty} dt \,
\, e^{i k^0 t} \, e^{-i k^d \epsilon \eta(t)} \;.
\end{equation}
Noting that $e^{-i k^d \epsilon \eta(t)}$ has a period $\tau = 4/\Omega$,
we expand it in Fourier space,
\begin{equation}
	e^{-i k^d \epsilon \eta(t)} \;=\; \sum_{n=-\infty}^{+\infty} C_n \,
	e^{- i \omega_n t} 
\end{equation}
with $\omega_n = \frac{2\pi n}{\tau}$, and
\begin{equation}
	C_n \;=\; \frac{1}{\tau} \; \int_0^\tau dt \, e^{-i k^d \epsilon
	\eta(t)} \,e^{i \omega_n t} \;. 
\end{equation}
 By means of the usual `uncertainty relation' between time and frequency,
 we note that the Fourier series above have an implicit cutoff:
 \begin{equation}
	 |\omega_n|_{\rm max} \;\sim\; (\Delta t)^{-1} \;,
 \end{equation}
 or:  
 \begin{equation}
	 |n_{\rm max}| \sim (\Omega \Delta t)^{-1} \;.
 \end{equation}

\begin{figure}[!ht]
\centering
\includegraphics[scale=1.0]{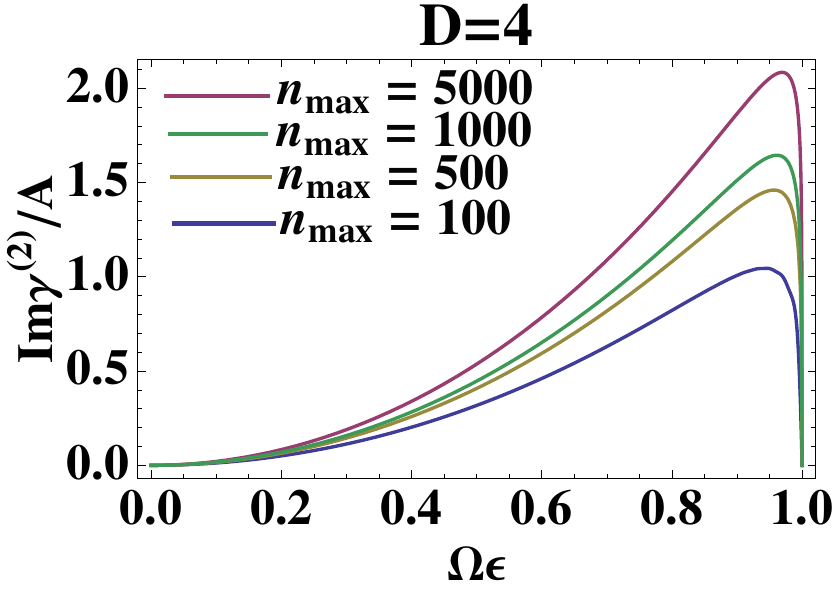} 
\caption{Imaginary part of the effective action  for a single mirror with
	relativistic motion, normalized by
	$A=\frac{\lambda^2\Omega}{64\pi^2}$, as a function of
	$\Omega\epsilon$. The different curves in the figure show the
	dependence with the returning time, i.e. with the maximum
acceleration of the mirror. } 
\label{fig:reld4} 
\end{figure}

The explicit form of the Fourier coefficients is: 
\begin{equation}
	\tilde{f}_\gamma(k^0,k^d) \;=\; 2 \pi \, \sqrt{1 -(\epsilon \Omega)^2} 
	\sum_{n = -\infty}^{+\infty} \,C_n \, \delta(k_0 - \omega_n) \;, 
\end{equation}
with:
\begin{equation}
	C_n \,=\, \frac{k^d \epsilon \Omega^2}{2 i} \, \frac{e^{- i k^d
	\epsilon} - (-1)^n \, e^{i k^d \epsilon} }{\omega_n^2 -
	(k^d \epsilon \Omega)^2}\;.	
\end{equation}
Then we see that the effective action per unit time and unit area, becomes: 
\begin{equation}\label{eq:gi22rsing}
	\gamma_I^{(2)} \;=\; \frac{1}{2}\, [1 - (\epsilon \Omega)^2 ] \, 
	\sum_{|n| < n_{\rm max}} \, 
	\int_{-\infty}^{+\infty} \frac{dk^d}{2\pi}\;
	\big[\widetilde{\Pi}^{(2)}(k)\big]\Big|_{k^0 = \omega_n ,\,{\mathbf k}_\parallel = {\mathbf 0}} \;|C_n|^2 \;,
\end{equation}
and its imaginary part in $D=4$ is:
\begin{align}\label{imgpertrel}
	{\rm Im}\Big[\gamma_I^{(2)}\Big]_{D=4} \,=&\, \frac{\lambda^2
\Omega}{64 \pi^2} \frac{1 - v^2}{v} \, 
\sum_{n=1}^{n_{\rm max}} \, \int_0^{\frac{n \pi}{2} v} du\;
	\frac{u^2}{[(\frac{n \pi}{2})^2 - u^2]^2} \\ \nonumber
	 &\times[1 - (-1)^n \cos( 2 u) ] \;.
\end{align}
By a change of variables in the integral it may be written as follows:
\begin{equation}\label{imgpertrel1}
	{\rm Im}\Big[\gamma_I^{(2)}\Big]_{D=4} \,=\, \frac{\lambda^2 \Omega}{32\pi^3}
	\frac{1 - v^2}{v} \, \int_0^v du\;
	\frac{u^2}{(1 - u^2)^2} \, g_4(u) \;, 
\end{equation}
with:
\begin{equation}\label{g4}
	g_4(u) \,=\, \sum_{n=1}^{n_{\rm max}} \; \frac{1 - (-1)^n \cos(n
	\pi u)}{n} \;. 
\end{equation}
This sum grows logarithmically with the cutoff.  
Fig. \ref{fig:reld4}  shows the results for different terms in the series, as a function
of the velocity. Recalling that $n_{max} \sim (\Omega \Delta t)^{-1}$, we
see that the imaginary part of the effective action grows as
$- \ln(\Omega \Delta t)$,  as $\Delta t \to 0$. 

It is worth observing that the dissipative effects vanish in the
ultrarelativistic limit. This is a consequence of the $\delta$-potential
interaction in our model (see Eq.\eqref{eq:single}), in which the effective
coupling between the mirror and
the quantum field is $\lambda/\gamma$: the mirror is transparent as $v\to 1$. 

It is interesting to repeat this calculation in $2+1$ dimensions, where
this kind of motion produces a finite imaginary part of the effective
action even in the limit of infinite acceleration.
The corresponding result is
\begin{equation}\label{imgpertrel2}
	{\rm Im}\Big[\gamma_I^{(2)}\Big]_{D=3} \,=\, \frac{\lambda^2}{4\pi^3}
	\frac{1 - v^2}{v} \, \int_0^v du\;
	\frac{u^2}{(1 - u^2)^2 \sqrt{1 - (\frac{u}{v})^2}} \, g_3(u) \;, 
\end{equation}
with:
\begin{equation}\label{g3}
g_3(u) \,=\, \sum_{n=1}^{+\infty} \; \frac{1 - (-1)^n \cos(n \pi u)}{n^2}=\frac{\pi^2}{4}(1-u^2) \;. 
\end{equation}
Inserting Eq.\eqref{g3} into Eq.\eqref{imgpertrel2} and performing the integral  we obtain
\begin{equation}
{\rm Im}\Big[\gamma_I^{(2)}\Big]_{D=3} \,=\, \frac{\lambda^2}{32}(1-v^2)\left[\frac{1}{\sqrt{1-v^2}}-1\right]\; .
\end{equation}

\begin{figure}[!ht]
\centering
\includegraphics[scale=1.0]{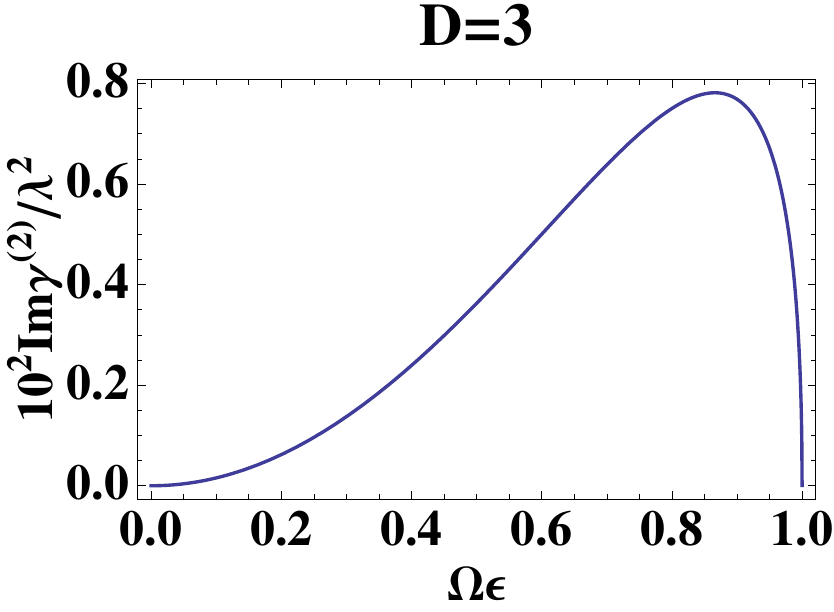} 
\caption{Imaginary part of the effective action (multiplied by $10^2$)  for a single mirror with relativistic motion, as a function of $v=\Omega\epsilon$, in $D=2+1$.
It is finite in the limit $\Delta t\to 0$, and has the same qualitative behavior than the case $D=3+1$}
\label{fig:reld3} 
\end{figure}

The plot in Fig. 3 shows the same qualitative behavior that in the previous case. The imaginary part of the effective action has a peak
at a velocity close to $v=0.9$ and then vanishes in the ultrarelativistic limit $v\to 1$.  

To conclude this section, we have seen that the imaginary part of the effective action diverges logarithmically with the maximum acceleration of the mirror in $3+1$ dimensions, and is finite in $2+1$ dimensions. It is also finite in $1+1$ dimensions \cite{Moore,FulDav}. 
The fact that the UV behavior of the imaginary part of the effective action get worse for higher dimensions is well known, and related to
the phase space available for the created particles.

\section{Two mirrors}
Let us consider now two mirrors, each one moving with a harmonic time
dependence, and having a constant average distance $a$:
\begin{equation}
q_L(t) \;=\; \epsilon_L \, \cos( \Omega_L t + \delta_L) \;,\;\;\; 
q_R(t) \;=\;  a \,+\, \epsilon_R \, \cos( \Omega_R t + \delta_R) \;.
\end{equation}
Note that we have included here also the phases $\delta_{L,R}$. As we will
see, they become relevant when there are two mirrors. 

The first relevant term in the perturbative expansion  corresponds again to
order-two. In this term, there are three contributions:
\begin{equation}
	\Gamma_I^{(2)} \;=\; \frac{i}{2} \, \langle ({\mathcal S}_L)^2
	\rangle_c
	\,+\, \frac{i}{2} \, \langle ({\mathcal S}_R)^2 \rangle_c
	\,+\, i \, \langle {\mathcal S}_L {\mathcal S}_R \rangle_c \;,
\end{equation}
the first two of which reduce to the second-order term for the respective
single mirror (the phases disappear in those terms). Thus, we will only
keep the last term, which is also the only contribution to this order which
depends on the average distance $a$. This term describes the `interference'
between mirrors in the particle creation rate.

Denoting that contribution by $\Gamma_{LR}^{(2)}$, we see that:
\begin{align}
	\Gamma_{LR}^{(2)} &=\; i \; \frac{\lambda_L}{2} \,
	\frac{\lambda_R}{2} \;
\int d^Dx \int d^Dx' \; \delta(z - q_L(t)) \,   2 \, \big( \langle
\varphi(x) \varphi(x') \rangle \big)^2 \delta(z' - q_R(t')) \nonumber\\
	& \equiv \; \frac{1}{2} \, 
\int d^Dx \int d^Dx' \; \delta(z - q_L(t)) \; \Pi_{LR}^{(2)}(x,x') \;
\delta(z' - q_R(t')) \;,
\end{align}
where
\begin{equation}
\Pi_{LR}^{(2)}(x,x') \;=\; i\lambda_L \, \lambda_R  \; \Big( G(x,x')
\Big)^2 \;,
\end{equation}
which is of course identical to the kernel $\Pi^{(2)}(x,x')$ except for a
multiplicative constant.

With a similar notation to the one used in the previous subsection, we also
see that:
\begin{equation}
	\Gamma_{LR}^{(2)} \;=\; \frac{1}{2}\, \int \frac{d^Dk}{(2\pi)^D}\; 
	\widetilde{\Pi}_{LR}^{(2)}(k) \, (\tilde{F}_L(k))^* \,  \tilde{F}_R(k) \;,
\end{equation}
where 
\begin{equation}
	\widetilde{\Pi}_{LR}^{(2)}(k) \;=\; i\lambda_L \, \lambda_R  \; \int
\frac{d^Dp}{(2\pi)^D} \; \widetilde{G}(p) \, \widetilde{G}(p-k)
\end{equation}
and:
\begin{align}
	& \tilde{F}_{L,R}(k) \;=\; \int d^Dx \, e^{i k \cdot x} \, \delta[x^d -
	q_{L,R}(x^0)] \,=\,(2\pi)^{d-1} \, \delta^{d-1}({\mathbf k}_\parallel) \, 
	\tilde{f}_{L,R}(k^0,k^d) \nonumber\\
	& \tilde{f}_{L,R}(k^0,k^d) \;=\; \int_{-\infty}^{+\infty} dt \, 
	e^{i k^0 t} \,  e^{-i k^d q_{L,R}(t)} \;.
\end{align}
Here:
\begin{equation}
	\tilde{f}_L(k^0,k^d) \;=\; 2\pi \, e^{i k^0
		\frac{\delta_L}{\Omega_L}} 
	\sum_{n=-\infty}^{+\infty} \,i^n \,  J_n(k^d \epsilon_L) \,
	\delta(k^0 - n \Omega_L) \;,
\end{equation}
and
\begin{equation}
	\tilde{f}_R(k^0,k^d) \;=\; 2\pi \, e^{i ( k^0
		\frac{\delta_R}{\Omega_R} - k^d a)} 
	\sum_{n=-\infty}^{+\infty} \,i^n \,  J_n(k^d \epsilon_R) \,
	\delta(k^0 - n \Omega_R) \;.
\end{equation}

Thus we see that the second order term will vanish unless $\Omega_L$ and
$\Omega_R$ satisfy:
\begin{equation}
	n_L \Omega_L = n_R \Omega_R
\end{equation}
for some natural numbers $n_{L,R}$, i.e., unless the frequencies are
commensurable. Among the different possibilities for this to happen, let us
first consider the simplest one, corresponding to equal frequencies:
$\Omega_L = \Omega_R \equiv \Omega$. Then,
\begin{align}
	{\rm Im}\Big[\gamma_{LR}^{(2)}\Big] &=\; \frac{1}{4\pi}\,
	\sum_{n=-\infty}^{+\infty} \, 
	\int_{-\infty}^{+\infty} dk^d\;
	\cos n \delta \; \cos k^d a  \nonumber\\
	&\times 
	{\rm Im}\big[\widetilde{\Pi}_{LR}^{(2)}(k)\big]\Big|_{k^0 = n \Omega,\,{\mathbf k}_\parallel = {\mathbf 0}} \; 
	J_n(k^d \epsilon_L) J_n(k^d \epsilon_R) 
\end{align}
with $\delta \equiv \delta_R - \delta_L$. Note that if we set $\delta = 0, \epsilon_L=\epsilon_R$ and $a=0$, the three contributions of $\Gamma_I^{(2)}$  
reduce to the result for a single mirror with coupling $\lambda_L+\lambda_R$, as expected.  

We will now discuss in some detail the particular case $a > 0, \epsilon_L=\epsilon_R=\epsilon$,  in four dimensions.
We have
\begin{equation}\label{2mirrorsexact}
	{\rm Im}\Big[\gamma_{LR}^{(2)}\Big] =\; \frac{\lambda^2}{16\pi^2 a}\,
	\sum_{n=1}^{+\infty} \, \cos n\delta\,	\int_{0}^{n\vert\Omega\vert a} dx\;
	  \cos x\,
	[J_n(x \epsilon/a)]^2 \, .
\end{equation}
As in the case of one mirror, we can evaluate the imaginary part of the effective action  perturbatively in the amplitude of the oscillation, keeping
only the $n=1$ term in the series, and expanding the Bessel function for small arguments. The result is
\begin{equation}\label{2mirrorspert}
	{\rm Im}\Big[\gamma_{LR}^{(2)}\Big] \simeq\;
	\frac{\lambda^2\epsilon^2}{64\pi^2 a^3} \, \left[2\Omega
		a\cos\Omega a+
	\left( (\Omega a)^2-2\right)\sin\Omega a\right]\, \cos \delta \; ,
\end{equation}	
and has some interesting properties. We first note that the sign of  ${\rm Im}\Big[\gamma_{LR}^{(2)}\Big]$ is not necessary positive, because this is only the 
interference part 
of the imaginary part of the effective action. It  vanishes for some particular values of $\Omega a=2.08, 5.94, ...$, etc, whatever the dephasing $\delta$. 
Moreover, it also vanishes for $\delta=\pi/2$, for any value of $a$.  In the particular case $\Omega a\ll 1$ we obtain
\begin{equation}
 {\rm Im}\Big[\gamma_{LR}^{(2)}\Big] \simeq \frac{\lambda^2\epsilon^2\vert\Omega\vert^3}{192 \pi^2}\cos\delta\, ,
\end{equation}
which is twice the result of a single mirror Eq.\eqref{imgpert} multiplied by $\cos\delta$.

\begin{figure}[!ht]
\centering
\begin{subfigure}[t]{.5\textwidth}
\centering
\includegraphics[width=\textwidth]{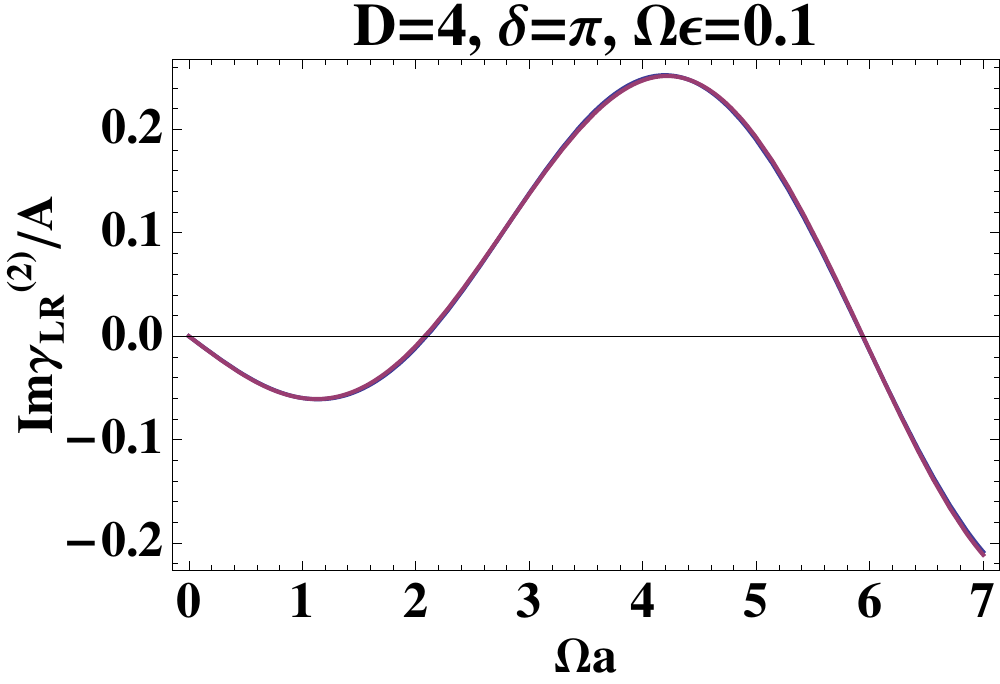}
\caption{} 
\label{fig2a}
\end{subfigure}~
\begin{subfigure}[t]{.5\textwidth}
\centering
\includegraphics[width=\textwidth]{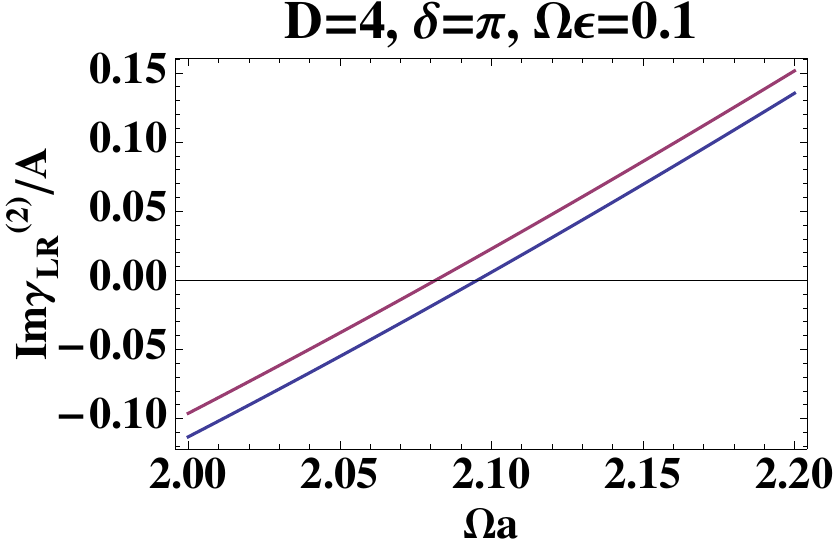} 
\caption{}
\label{fig2b}
\end{subfigure}
\caption{Imaginary part of the effective action for two mirrors (normalized with $A=\frac{\lambda^2}{64\pi^2 a}\times 10^{-2}$ for (\subref{fig2a}) and $A=\frac{\lambda^2}{64\pi^2 a}\times 10^{-3}$ for (\subref{fig2b})), as a function of $\Omega a$. (\subref{fig2a}) The exact (red) and perturbative (blue) results are almost indistinguishable. (\subref{fig2b}) Zoom  near a zero of the imaginary part of the effective action.}
\end{figure}

\begin{figure}[!ht]
\centering
\begin{subfigure}[t]{.5\textwidth}
\centering
\includegraphics[width=\textwidth]{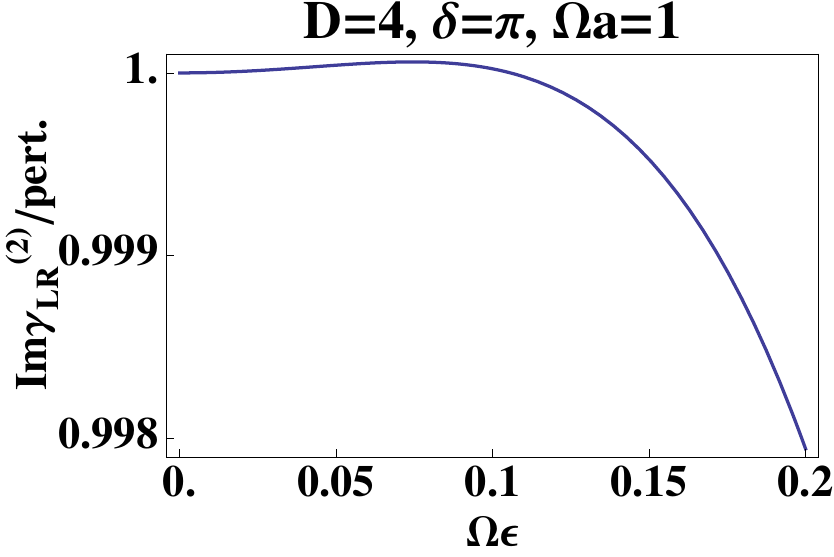}
\caption{} 
\label{fig3a}
\end{subfigure}~
\begin{subfigure}[t]{.5\textwidth}
\centering
\includegraphics[width=\textwidth]{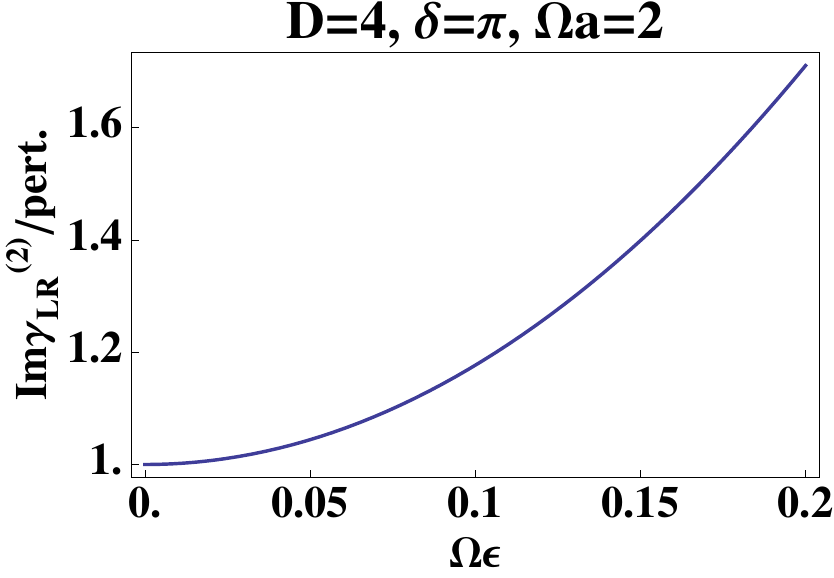} 
\caption{}
\label{fig3b}
\end{subfigure}
\caption{Ratio of the exact and perturbative results as a function of $\Omega\epsilon$, for (\subref{fig3a}) $\Omega a=1$ and (\subref{fig3b}) $\Omega a=2$. In the case of $\Omega a=2$ the figure illustrates the difference between perturbative and exact results in a region close to a zero of the imaginary part of  the efective action.}
\end{figure}

It is interesting to asses the accuracy of the perturbative result Eq.\eqref{2mirrorspert}. In Fig \ref{fig2a} we plot ${\rm Im}\Big[\gamma_{LR}^{(2)}\Big]$ for a fixed value of
$\Omega\epsilon$, as a function of $\Omega a$. The exact Eq.\eqref{2mirrorsexact} and perturbative Eq.\eqref{2mirrorspert} results are indistinguishable, 
unless $\Omega a$ is close to a zero of the perturbative result. This is illustrated in Fig. \ref{fig2b}. The terms of higher order produce only a small shift in the position of 
the zeros.\\
In Figs. \ref{fig3a} and \ref{fig3b} we plot the ratio of the exact and perturbative results as a function of $\Omega\epsilon$, for fixed values of $\Omega a$. These results verifies the significant difference between perturbative and exact results in the region near to zeros of the imaginary part of the effective action.\\

\section{Higher orders, and more mirrors}

The results presented in the previous sections can be easily generalized to include higher order corrections
in $\lambda$. For example, the third order contribution to the effective action for a single, non-relativistic mirror,  reads 
\begin{eqnarray}
	\Gamma_I^{(3)}=\frac{1}{3!} \int d^Dx_1 \int d^Dx_2 && \int 
	d^Dx_3  \Pi^{(3)}(x_1,x_2,x_3)  \delta(z_1 - q(t_1)) \, 
	\nonumber\\ && \times  \delta(z_2
	- q(t_2))  
	 \delta(z_3 - q(t_3)) \, ,
\end{eqnarray}
where
\begin{equation}
	\Pi^{(3)}(x_1,x_2,x_3) \;=\; \frac{1}{2} \, \lambda^3 \; G(x_1 -
	x_2) \,G(x_2 - x_3) G(x_3 - x_1) \; .
\end{equation}
Now, since $\Pi^{(3)}$ depends only on the differences between pairs of
arguments, its Fourier transform may be written as follows:
\begin{equation}
	\widetilde{\Pi}^{(3)}(k_1,k_2,k_3) \;=\; (2\pi)^D
	\delta(k_1+k_2+k_3) \;\widetilde{\Pi}^{(3)}(k_1,k_2)  \; .
\end{equation}
Then, the effective action becomes extensive in time and the parallel
coordinates; defining a density as in the second-order case:
\begin{align}\label{eq:gi3sing}
	\gamma_I^{(3)} &=\; \frac{1}{3!}\, \sum_{n_1,
	n_2=-\infty}^{+\infty} \, 
	\int \frac{dk_1^d}{2\pi} \frac{dk_2^d}{2\pi}\;
	\widetilde{\Pi}^{(3)}(n_1 \Omega, {\mathbf 0}_\parallel, k_1^d; n_2
	\Omega, {\mathbf 0}_\parallel, k_2^d) \nonumber\\ 
	&\times	J_{n_1}(k_1^d \epsilon) \,
		J_{n_2}(k_2^d \epsilon) \,
		J_{-n_1-n_2}(-(k_1^d+k_2^d) \epsilon) \;.
\end{align}
Again, the imaginary part is determined by the corresponding absorptive
part in $\widetilde{\Pi}^{(3)}$, that is given by
\begin{align}
& \widetilde{\Pi}^{(3)}(k_1,k_2) \;=\;
\frac{\lambda^3}{(4\pi)^{D/2}} \, \Gamma(3-D/2) \,
\int_0^1 d\alpha_1 \int_0^1 d\alpha_2 \;\theta(1-\alpha_1-\alpha_2)\,
\nonumber\\
&\times \; \Big[ 
\big(\alpha_1 k_1 + \alpha_2 (k_1 + k_2) \big)^2 - \alpha_1 k_1^2 - \alpha_2
(k_1+k_2)^2 \Big]^{D/2 - 3}\;.
\end{align}

More generally, the term of order $n$
can be written as
follows
\begin{align}\label{eq:gin}
\Gamma_I^{(n)} &=\; \frac{1}{n!} \, 
\int d^Dx_1 \int d^Dx_2 \ldots \int d^Dx_n \;\Pi(x_1,x_2,\ldots,x_n) \nonumber\\& \times \,\,\delta(z_{1} - q(t_{1})) \ldots \,
\delta(z_n - q(t_n)) \;, 
\end{align}
where:
\begin{align}
	\Pi(x_1,x_2,\ldots,x_n) &\;=\; \frac{i^{n-1}}{2}\,(-1)^n \,
\lambda^n \; G(x_1-x_2)  \nonumber\\ 
&\;\times\;G(x_2-x_3) \ldots G(x_n-x_1) \;,
\end{align}
of which  only the part of which is completely symmetric with respect to
its arguments contributes to Eq.(\ref{eq:gin}).

When considering $N$ mirrors, the interaction action ${\mathcal
S}_I $ will be of the form
\begin{equation}
{\mathcal S}_I=\sum_{i=1}^N  {\mathcal S}_I^{(i)}\, ,
\end{equation}
where each term is proportional to a coupling constant $\lambda^{(i)}$. The contribution of order $n$ to the effective action will be proportional to 
$\langle ({\mathcal S}_I)^n \rangle_c$. Therefore, we see that the expansion
in powers of the coupling constants $\lambda^{(i)}$ (assuming all of them to be of the same order)
includes, up to order $n<N$,  $n-$body interactions between the mirrors. 
This is entirely analogous to
what happens for the static Casimir energy, when computed using 	
a perturbative expansion in the dielectric contrast \cite{Golestanian}.

\section{Conclusions}

We have considered the DCE for semitransparent
mirrors, having as the main goal the development of a systematic
perturbative approach to compute the imaginary part of the effective action
in powers of the coupling constant between the mirrors and the quantum
field. The approach is valid in dimensions $D\geq 3$, because of the infrared
divergences that arise in $D=2$.

We presented explicit results for the case of one or two oscillating
mirrors, considering both relativistic and non-relativistic motions, without
restricting the calculations to the small amplitude limit.  Technically, this 
has been accomplished by using the Jacobi-Anger expansion. The resulting expressions
for the imaginary part of the effective action are suitable to compute relativistic corrections,
expanding the $\gamma^{-1}$ factors in powers of $v^2$. 

For the case of a single mirror undergoing harmonic oscillations, we have
shown that the results in the non-relativistic case are practically
identical to the ones of a perturbative expansion in the amplitude of
oscillation, to the lowest non trivial order.  However, in our approach is very simple to incorporate
relativistic corrections, going beyond the usual non-relativistic  results. 
Then we considered, for the
case of a single mirror, an example of ultrarelativistic motion which
corresponded to accelerations concentrated in time at the return points,
having an essentially constant speed elsewhere. For $2+1$ and $3+1$
spacetime dimensions, we found a qualitatively similar behaviour, where the
dissipation reaches a maximum at a relativistic speed, but vanishes when
the speed of the oscillation reaches the speed of light. In $2+1$ dimensions the
imaginary part of the effective action is finite in the limit of infinite acceleration, while in $3+1$
dimensions diverges logarithmically. 

The case of two mirrors oscillating about a constant distance $a$ is
qualitatively different. Indeed, we have found departures from the
calculation perturbative in the amplitudes in the non-relativistic
case (the only case we considered for two mirrors). Those departures are
concentrated close to the zeroes of the imaginary part of the effective
action, regarded as a function of $\Omega a$. Moreover, we have found that
the interference term in the effective action vanishes unless the mirrors move with
commensurable frequencies.

Finally, we considered briefly the calculation of higher order corrections, for an arbitrary number of mirrors.
It is conceptually interesting to remark that the contribution of order $n$ to the effective action includes
up to $n$-body interactions, as happens when computing the static Casimir energy for dilute
bodies \cite{Golestanian, Milton}

Our approach can be generalized to more realistic situations involving the
quantum electromagnetic field where the `mirrors' are, for instance,
refractive index perturbations, or relativistic flying mirrors in a plasma \cite{RFM}.  
Moreover, it is also possible to consider other geometries for the mirrors.

\section*{Acknowledgements} This work was supported by CONICET, ANPCyT and UNCuyo.



\begin{thebibliography}{bib}
\bibitem{footnote}We use
`planes' to denote flat spatial regions with codimension $1$ relative to
$d$-dimensional space. They are planes in the usual sense only when
$d=3$.
\end{thebibliography}
\end{document}